\documentstyle{mn}

\newif\ifAMStwofonts
\ifoldfss
  \ifCUPmtlplainloaded \else
    \NewTextAlphabet{textbfit} {cmbxti10} {}
    \NewTextAlphabet{textbfss} {cmssbx10} {}
    \NewMathAlphabet{mathbfit} {cmbxti10} {} % for math mode
    \NewMathAlphabet{mathbfss} {cmssbx10} {} %  "   "    "
  \fi
  \ifAMStwofonts
    \ifCUPmtlplainloaded \else
      \NewSymbolFont{upmath} {eurm10}
      \NewSymbolFont{AMSa} {msam10}
      \NewMathSymbol{\upi}     {0}{upmath}{19}
      \NewMathSymbol{\umu}     {0}{upmath}{16}
      \NewMathSymbol{\upartial}{0}{upmath}{40}
      \NewMathSymbol{\leqslant}{3}{AMSa}{36}
      \NewMathSymbol{\geqslant}{3}{AMSa}{3E}

       \let\le=\leqslant
       
    \fi
  \fi
\fi % End of OFSS

\ifnfssone
  \newmathalphabet{\mathit}
  \addtoversion{normal}{\mathit}{cmr}{m}{it}
  \addtoversion{bold}{\mathit}{cmr}{bx}{it}
  \newmathalphabet{\mathbfit} % math mode version of \textbfit{..}
  \addtoversion{normal}{\mathbfit}{cmr}{bx}{it}
  \addtoversion{bold}{\mathbfit}{cmr}{bx}{it}
  \newmathalphabet{\mathbfss} % math mode version of \textbfss{..}
  \addtoversion{normal}{\mathbfss}{cmss}{bx}{n}
  \addtoversion{bold}{\mathbfss}{cmss}{bx}{n}
  \ifAMStwofonts
    \ifCUPmtlplainloaded \else
      \UseAMStwoboldmath
      \makeatletter
      \new@mathgroup\upmath@group
      \define@mathgroup\mv@normal\upmath@group{eur}{m}{n}
      \define@mathgroup\mv@bold\upmath@group{eur}{b}{n}
      \edef\UPM{\hexnumber\upmath@group}
      \new@mathgroup\amsa@group
      \define@mathgroup\mv@normal\amsa@group{msa}{m}{n}
      \define@mathgroup\mv@bold\amsa@group{msa}{m}{n}
      \edef\AMSa{\hexnumber\amsa@group}
      \makeatother
      \mathchardef\upi="0\UPM19
      \mathchardef\umu="0\UPM16
      \mathchardef\upartial="0\UPM40
      \mathchardef\leqslant="3\AMSa36
      \mathchardef\geqslant="3\AMSa3E

       \let\le=\leqslant

    \fi
  \fi
\fi % End of NFSS release 1

\ifnfsstwo
  \DeclareMathAlphabet{\mathbfit}{OT1}{cmr}{bx}{it}
  \SetMathAlphabet\mathbfit{bold}{OT1}{cmr}{bx}{it}
  \DeclareMathAlphabet{\mathbfss}{OT1}{cmss}{bx}{n}
  \SetMathAlphabet\mathbfss{bold}{OT1}{cmss}{bx}{n}
  \ifAMStwofonts
    \ifCUPmtlplainloaded \else
      \DeclareSymbolFont{UPM}{U}{eur}{m}{n}
      \SetSymbolFont{UPM}{bold}{U}{eur}{b}{n}
      \DeclareSymbolFont{AMSa}{U}{msa}{m}{n}
      \DeclareMathSymbol{\upi}{0}{UPM}{"19}
      \DeclareMathSymbol{\umu}{0}{UPM}{"16}
      \DeclareMathSymbol{\upartial}{0}{UPM}{"40}
      \DeclareMathSymbol{\leqslant}{3}{AMSa}{"36}
      \DeclareMathSymbol{\geqslant}{3}{AMSa}{"3E}

       \let\le=\leqslant

    \fi
  \fi
\fi % End of NFSS release 2

\ifCUPmtlplainloaded \else
  \ifAMStwofonts \else % If no AMS fonts
    \def\upi{\pi}
    \def\umu{\mu}
    \def\upartial{\partial}
  \fi
\fi

\title[Optical counterparts of microJansky radiosources]
{The Canada-France Redshift Survey VII:\\ Optical counterparts of microJansky
radiosources}

\author[F. Hammer et al.]
       {F.~Hammer,$^1$\thanks{Visiting Astronomer, Canada-France-Hawaii
Telescope,
which is operated by the National Research Council of Canada, the Centre de
Recherche Scientifique of France, and the University of Hawaii}
        David Crampton,$^{2\star}$
        Simon J.~Lilly,$^{3\star}$
        O.~Le F\`evre,$^{1\star}$
        T.~Kenet$^3$\\
       $^1$ DAEC, Observatoire de Meudon, Meudon, 92195, France\\
       $^2$ Dominion Astrophysical Observatory, National Research Council of
Canada, Victoria, V8X 4M6 Canada\\
       $^3$ Department of Astronomy, University of Toronto, Toronto, M5S 1A7
Canada}
\date{Accepted .
      Received 1994 December }

\pagerange{\pageref{firstpage}--\pageref{lastpage}}
\pubyear{1995}

\begin{document}

\maketitle

\label{firstpage}

\begin{abstract}
Deep imaging and spectroscopy have been carried out for optical counterparts of
a complete sample of S$>$16 $\mu$Jy radiosources during the course of the
Canada-France Redshift Survey (CFRS). All 36 sources but two
have been optically identified, and spectra have been obtained
for 23 of them. The objects brighter than $I_{AB}<$22.5 for which
 we have spectra reveal
three populations dominating the $\mu$Jy
radio counts: z$>$0.7 early-type galaxies with radio emission powered by an
AGN, intermediate redshift post-starburst galaxies, and
lower redshift blue emission-line objects. From their radio and optical
properties, it is argued that the 11 objects fainter than $I_{AB}>$22.5 are
mostly at  z$>$1, and one half of them are probably early-type galaxies.
We conclude that $\sim$40 per cent of the $\mu$Jy sources are likely to be at
z $>$1.

Between one third and one half of the luminous ellipticals in this field
beyond z = 0.7 show
moderately powerful radio emission (P $\sim$ 5
10$^{23}$ W Hz$^{-1}$)  which is at least 10 times
more powerful than seen in local samples, and probably reflects
 evolution of the activity in their nuclei. Only
one classical starburst galaxy is identified in the sample;
the rest of the blue emission-line
objects show optical and radio activity more typical of low power AGNs
than starbursts. The
number of post-starbursts at $\mu$Jy levels is considerably higher than
 the surface
density of mJy starburst galaxies, suggesting the latter are the parent
population of the former. While starburst galaxies are considered to be
major contributors to
the mJy radiosources counts, the majority  of the $\mu$Jy radio sources
appear to be related to AGN activity rather than to normal star formation.

\end{abstract}


\section{Introduction}
Deep 1.4GHz counts show an upturn below a few milliJanskys (mJy),
corresponding to a rapid increase in the number of faint sources
(Windhorst, 1984; Condon and Mitchell, 1984; Oort and Windhorst, 1985),
with a decreasing fraction of sources which can be ascribed to
elliptical galaxies and QSOs. Several scenarios
have been developed to interpret this numerous new population, invoking
either a non-evolving population of low redshift galaxies at very low radio
power (Wall et al. 1986) or strong evolution of normal spirals at z$>$0.1
(Condon 1989). The latter scenario is supported by the spectroscopic work of
Benn et al. (1993), who identified the brightest optical counterparts of
sub-mJy radiosources as z$\sim$0.2 blue galaxies which they
interpreted as starburst galaxies.
However, there is little information about the true redshift distribution of
the sub-mJy population because only a small fraction of them have been
identified and have measured redshifts (Rowan-Robinson et al. 1993).
Towards the $\mu$Jy levels ($\sim$ 16$\mu$Jy at 4.86GHz), the
projected number density of
radiosources reaches the number density of B = 21.5 field galaxies, while the
fraction of flat spectrum radiosources is increasing continuously (Fomalont
et al. 1991, hereafter FWKK). Because of this high surface density, the
understanding of $\mu$Jy radiosources may provide new and interesting
constraints on the evolution of field galaxies.

During the preliminary deep imaging phase
of our large spectroscopic survey of faint field galaxies (CFRS),
one of our fields was chosen to coincide with the FWKK
radiosource field (see Tresse et al. 1993). This field -- as with
other CFRS fields -- is at high Galactic latitude
 (b$^{II}$ $\sim$ 60$^{\circ}$).
We present here the results of the observations of 36 objects which
represent a complete sample of
$\mu$Jy radiosources. Although we were only able to
acquire spectra for objects with
I$_{AB} \le $22.5, the colours and radio spectral indices of the fainter
objects enable us to investigate the properties of the entire radio sample.

\section{Optical identification, photometry, and spectroscopy}

\subsection{Photometry and optical morphology}

The field at 1415+52 is one of the five 10\arcmin $\times$  10\arcmin
{}~ fields selected for the CFRS survey of I$_{AB}$ $\le$22.5 galaxies.
BVI photometry was carried out at
CFHT with FOCAM, to a depth well beyond the spectroscopic limits to ensure that
there was no bias against low surface brightness objects.
The image quality was generally excellent,
 ranging from FWHM  = 0\farcs7 -- 0\farcs9.
$K^{\prime}$ band photometry was also
obtained at CFHT with RedEye on roughly one-third of the spectroscopic objects.
 Details of the data reduction and
photometric limits are presented in  CFRS I: Lilly et al. 1994. Throughout
this paper the AB system is used, where V$_{AB}$ = V, I$_{AB}$ = I + 0.48,
K$_{AB}$ = K + 1.78.

The depth and quality of our images also allows us to examine the
morphology of all of the optical counterparts. The CCD pixel sizes
of the B, V and I  images are 0\farcs207 and the FWHM of point sources
are typically 2.4 pixels after deconvolution. Sources were systematically
deconvolved in the I frame, and very often in the V frame, using
the {\sc MEM} tool of the {\sc IRAF} {\sc STSDAS} package and
a relatively bright star selected as close as
possible to the source.

\subsection{Astrometry}

In order to derive very accurate astrometric positions for the objects on
our FOCAM images, accurate positions of several secondary standard
stars in the field were utilized. To accomplish this,
accurate positions of 18 HST Guide Star Catalogue (GSC)
 stars were measured with a two-coordinate measuring machine on POSS E plates
in an area which included the CCD field, plus the positions of
16 stars which were unsaturated on the CCD.
An astrometric plate solution for the GSC star positions was
then used to relate pixel positions of the 16 stars on the CCD
image to their astrometric positions, with a formal RMS error 0\farcs46,
and to develop a similarly precise
mapping of pixel positions to celestial coordinates for
all objects in the field. As will become apparent
below, the resulting positions for the optical counterparts are in
excellent agreement with the radio source positions.

FWKK list 62 sources in their complete (S $>$ 16 $\mu$Jy) sample;
 40 of these lie in our
survey field. One source (15V56) has to be excluded from our subsample
since it lies in the
outskirts of a bright nearby galaxy,
preventing optical identification to any reasonable depth
(as noted below, others were subsequently
 deleted for other reasons, reducing the number to 36).
We have been very successful in identifying virtually all of the
remaining sources with optical counterparts on our images.
Comparison of the optical and radio positions of 19 sources which
are isolated (no confusing adjacent sources), and which are obvious,
certain identifications, demonstrates that the optical
and radio reference frames are in perfect agreement (to better than 0\farcs15).
The random error, as indicated by the RMS residual in the optical and
radio positions for these
 sources, is 0\farcs54.

\subsection{Spectroscopy}

Spectroscopic observations of 11 of the counterparts were
obtained with the CFHT MOS/SIS multi-object spectrographs
as part of the CFRS survey, and subsequently an
additional 12 were specifically targeted in order to achieve reasonable
completeness. Including a few objects observed by Tresse et al. (1993),
spectra of 22 of the 25 counterparts
in the double-limited sample (S $>$16$\mu$Jy and I$_{AB}$ $\le$22.5)
were obtained.
The spectral resolution was 40\AA, and integration times ranged from 7$^{h}$ to
9$^{h}$ (7 to 9 times one hour integration).
The methodology adopted for data
reduction, redshift identification and reliability is discussed in
detail in CFRS II: Le F\`evre et al. (1995), and data for other objects
in this field are given in CFRS III: Lilly et al. (1995).

\subsection{Identifications}

A summary of the results of our identifications, photometry and
spectroscopy for all 39 sources is given in Table 1.  FWKK's source
number and radio (5 GHz) flux are listed in the first two columns, followed by
the
number of the optical counterpart in the CFRS catalogue.
The isophotal I$_{AB}$ magnitude is listed in column 4, followed by
 I, V and K  three-arcsec aperture magnitudes, measured redshifts, `galaxy
type',
radio spectral
indices and \hbox{[O\,{\sc ii}]} (rest) equivalent widths. The `galaxy types'
given in the table represent broad divisions on the basis of their spectra,
colour indices and morphologies (see below)
 into ellipticals (E), post-starburst
(S+A -- see section 3.3), and emission-line galaxies (EM). Question marks
appended to these `types' indicate that the classification was based on colours
and
radio properties.
Note that several of the sources in the FWKK `complete' sample
(with 8 arcsec resolution)
 are actually  blends of sources which are resolved at their highest
 resolution (2\farcs5, 4.8GHz).
  In most cases,
each of these individual radiosources were also identified with optical
counterparts. However, since some of these resolved sources
have no individual sources with flux higher than 16$\mu$Jy,
they should not be included in the complete subsample. The five
sources listed at the bottom of Table 1 are in
this category.
After these sources have been removed (and one double source counted
as two), the complete sample in our
MOS field is reduced to 36 radio sources, almost all of which
are unresolved, even at  FWKK's highest resolution.
Only two radio sources (15V62 and 70) remain unidentified, even in the
sum of our deep optical images from the B to the K band.
The resulting completeness in optical identification is 94 per cent.

The positional offsets
and discussions of the identifications of the objects are given in the notes
to Table 1, and the radio positions are also indicated on images of all the
40 sources that lie in the field in Fig. 1.

\begin{table*}
\begin{minipage}{120mm}
 \caption{Optical Counterparts of $\mu$Jy Sources}
 \label{symbols}
\ Objects with $S>16\mu$Jy and $I_{AB}\le$22.5
\smallskip

 \begin{tabular}{crcccccccrccc}
FWKK & S & CFRS & $I_{AB}$ & I(3") & V(3") & K(3") & z & Type & $\alpha$ & err
& $W_{[OII]}$ & err\\
15V..& $\mu$Jy & 14..& mag & mag & mag & mag &&&&& \AA & \AA \\
 \hline
5 & 41 & 1501 & 21.74 & 22.04 & 23.16 &&& S+A? & 0.2 & 0.3 &&\\
10 & 1912 & 1373 & 21.77 & 22.27 & 24.55 & 20.56 && E? & $<$--0.2 & 0.1 &&\\
11 & 70 & 1329 & 19.49 & 19.86 & 20.94 && 0.375 & S+A & 0.5 & 0.1 & 15 & 5\\
12 & 45 & 1303 & 19.97 & 20.14 & 20.05 & 19.25 & 0.985 & QSO & 0.1 & 0.3 & 30 &
3\\
15 & 24 & 1246 & 22.18 & 22.30 & 23.92 &&  & E? & $<$--0.3 &&&\\
19 & 24 & 1190 & 20.99 & 21.17 & 22.67 && 0.754 & S+A & 0.7 & 0.3 & 10 & 3\\
21 & 298 & 1177 & 20.78 & 21.29 & 22.68 && 0.724 & S+A & 0.3 & 0.1 & 68 & 20\\
23 & 54 & 1157 & 20.54 & 21.08 & 22.99 && 1.149 & ? & 0.3 & 0.2 & 33 & 10\\
24 & 79 & 1139 & 20.20 & 20.64 & 21.93 & 18.92 & 0.660 & S+A & 0.4 & 0.1 & 16 &
4\\
26a & 47 & 1041 & 21.7 & 21.87 & 22.36 && 0.372 & EM & $<$--0.1 && 75 & 7\\
26b & 47 & 9025 & 18.31 & 19.27 & 20.04 && 0.155 & EM & $<$--0.1 &&&\\
28 & 31 & 1028 & 21.57 & 21.74 & 24.01 & 19.69 & 0.988 & E & --0.3 & 0.6 & 31 &
15\\
34 & 1311 & 0937 & 21.41 & 22.01 & 24.42 && 0.838 & E & $<$--0.2 && 12 & 8\\
39 & 35 & 0854 & 21.7 & 21.85 & 24.08 & 19.59 & 0.992 & E & --0.8 & 0.7 &&\\
40 & 33 & 0820 & 21.69 & 21.94 & 24.52 & 19.37 & 0.976 & E & --0.1 & 0.4 & 24 &
12\\
42 & 14 & 0727 & 20.62 & 21.05 & 21.90 && 0.463 & EM &&& 40 & 8\\
47 & 53 & 0665 & 22.41 & 23.00 & 23.67 & 21.93 && ? & 0.5 & 0.2 &&\\
48 & 20 & 0663 & 20.88 & 21.11 & 22.57 && 0.743 & S+A & 0.2 & 0.4 & 11 & 4\\
49 & 31 & 0667 & 19.48 & 19.96 & 20.65 & 18.92 && ? & 0.7 & 0.3 &&\\
50 & 705 & 0645 & 22.44 & 22.76 & 24.72 &&& ? & 0.7 & 0.1 &&\\
57 & 37 & 0573 & 16.90 & 17.70 & 17.89 & 17.53 & 0.010 & EM & $<$--0.5 &&&\\
60 & 39 & 9154 & 21.57 & 21.94 & 23.43 && 0.812 & S+A & 0.3 & 0.2 & 75 & 30\\
65 & 35 & 0426 & 19.80 & 19.86 & 21.48 && 0 & M$\star$ & 0.4 & 0.4 &&\\
73 & 37 & 0276 & 20.67 & 20.95 & 22.93 && 0.746 & E & $<$--0.5 &&&\\
81 & 28 & 0154 & 22.08 & 22.16 & 22.74 && 1.158 & EM & 0.7 & 0.2 & 55 & 10\\
 \end{tabular}
 \medskip

\ Objects with $S>16\mu$Jy and $I_{AB}>$22.5
\smallskip

\begin{tabular}{rrrllllccrlrr}
FWKK & S & CFRS & I$_{AB}$ & I(3") & V(3") & K(3") & z & Type & $\alpha$ & err
& $W_{[OII]}$  & err\\
15V..& $\mu$Jy & 14..& mag & mag & mag & mag &&&&& \AA & \AA \\
 \hline
18 & 44 &&& $>25$ & $>26$ & \ \ \ 20.8 &&& 0.2 & 0.3 &&\\
33 & 20 & 9991 & 23 & \ \ \ 23.2 & \ \ \ 24.7 &&& E? & $<$--0.2 &&&\\
37 & 51 & 0861 & 23.5 & \ \ \ 23.65 & \ \ \ 25.25 & \ \ \ 20.23 && E? & --0.4 &
0.4 &&\\
45 & 33 & 9993 & 24.1 & \ \ \ 24.3 & $>26$ &&& E? & --0.2 & 0.7 &&\\
51 & 41 & 9994 & 22.99 & \ \ \ 23.14 & $>26$ &&&& 0.4 & 0.2 &&\\
53 & 30 & 0612 & 23.0 & \ \ \ 23.04 & \ \ \ 23.95 &&& S+A? & 0.7 & 0.2 &&\\
59 & 19 & 9995 & 24.3 & \ \ \ 24.5 & \ \ \ 24.4 &&&& 0.1 & 0.5 &&\\
62 & 24 &&& $>25$ & $>26$ & $>21$ &&& $<$--0.3 &&&\\
67 & 46 &&& $>25$ & $>26$ &&&& --0.3 & 0.3 &&\\
70 & 576 &&& $>25$ & $>26$ &&&& 0.6 & 0.1 &&\\
72 & 32 & 9996 & 23.6 & \ \ \ 23.8 & $>26$ & \ \ \ 20.6 && E? & --0.5 & 0.5
&&\\
\end{tabular}

\medskip

\ Objects with $S<16\mu$Jy after deblending of the radio emission

\begin{tabular}{rrrlllllclclr}
\smallskip
FWKK & S & CFRS & I$_{AB}$ & I(3") & V(3") & K(3") & z & type & $\alpha$ & err
& $W_{[OII]}$  & err\\
15V..& $\mu$Jy & 14..& mag & mag & mag & mag &&&&& \AA & \AA \\
 \hline
30 & $<26$ & 0983 & 21.27 & 21.53 & 21.84 & 20.79 & 0.286 & EM & \ 0.9? & 0.2 &
58 & 15\\
30 & $<26$ & 0998 & 20.58 & 20.73 & 21.99 & 19.06 & 0.43 &  & \ 0.9? & 0.2 & 18
& 4\\
36 & $<22$ & 0887 & 20.46 & 20.69 & 22.16 &&&& \ 0.2 & 0.5 &&\\
41 & $<38$ & 0818 & 21.02 & 21.33 & 22.45 && 0.899 &  & --0.4 & 0.8 & 39 & 10\\
69 & 13 & 0393 & 20.44 & 20.66 & 21.64 & 19.76 & 0.602 &  & --0.3 & 0.5 & 36 &
8\\
\end{tabular}

\end{minipage}

\end{table*}

\subsection{Notes on individual sources}

\noindent
15V5: Excellent positional agreement (0\farcs1) with CFRS\-14.1501 but
 no spectrum was obtained.

\noindent
15V10: CFRS14.1373 is slightly offset (1\farcs6) from the radio core
of this extended double source as shown in the map by FWKK,
    but it appears to be the probable identification.
    An uncertain redshift of 0.652  was derived, but no \hbox{[O\,{\sc ii}]}
emission,
    characteristic of most of these radio sources, is detected.

\noindent
15V11: Excellent positional agreement (0\farcs5) with CFRS\-14.1329.

\noindent
15V12: Excellent positional agreement (0\farcs4) with CFRS\-14.1303, a quasar
at
   z = 0.985 associated with a large structure of galaxies (Le F\`evre et al.
   1994) in this field.

\noindent
15V15: Only fair positional agreement (2\farcs0) with  CFRS\-14.1246
 for which we derive a tentative redshift
   of z = 0.65.  FWKK claim it is ``at edge of galaxy with
   a double nucleus'', but the faint objects shown in Fig. 1 appear to be
   isolated galaxies.

\noindent
15V18: Nothing visible at the radio source position on the
 summed B, V and I image, but a faint counterpart is visible on the K image.
It is thus extremely red, $(I - K)_{AB} >$ 4.4, and probably at z $>$ 1.

\noindent
15V19: Excellent positional agreement (0\farcs2) with CFRS\-14.1190.

\noindent
15V21: Poor positional agreement (4\farcs6) with CFRS\-14.1177 which has z =
0.724.
   Although the spectrum of this galaxy shows \hbox{[O\,{\sc ii}]} emission,
typical of the
   optical counterparts, it appears that the fainter (I$_{AB} \sim$ 23.9)
galaxy
   which is only 2\farcs0 away is a more likely counterpart. FWKK
   comment that a source 13 arcsec E confused the map of this source, possibly
   the explanation of the poor positional agreement. Unfortunately, no
   spectrum of the fainter galaxy was obtained.

\noindent
15V23: Excellent positional agreement (0\farcs4) with CFRS\-14.1157, a galaxy
with
a peculiar morphology and a peculiar
red spectrum displaying strong emission at 8008\AA\ identified
   as 3727, and possibly other weaker emission lines at this redshift.
   We adopted z = 1.149 as the most likely redshift,
   but z = 0.372 cannot be ruled out. If it is at the high redshift, it
is similar in some respects to the well-known powerful radiosources at similar
redshifts.

\noindent
15V24: Excellent positional agreement (0\farcs6) with CFRS\-14.1139.

\noindent
15V26: According to FWKK, the radio emission comes from both
   the bright spiral galaxy (CFRS14.9025) and the fainter galaxy (CFRS14.1041)
   to the east. The latter object
   shows strong \hbox{[O\,{\sc ii}]} and \hbox{[O\,{\sc iii}]} emission lines;
the brighter one
   shows strong H$\alpha$ and [SII] emission.

\noindent
15V28: Excellent positional agreement (0\farcs7) with CFRS\-14.1028.
FWKK give 31$\mu$Jy for the radio
   emission of the ``core''
   associated with this galaxy, and comment that the extended emission
   (see their map) may come from other galaxies.

\noindent
15V30: There are two radio lobes, each centered on a galaxy; the northern
   one (CFRS14.983) is a strong emission line object at z = 0.286,
   and the other (CFRS14.0998) has a more normal spectrum.

\noindent
15V33: The radio source appears to be associated with a faint red galaxy.

\noindent
15V34: There is excellent positional agreement (0\farcs3) of this strong radio
source with  CFRS14.0937.

\noindent
15V36: Although the position of this radio source is 3\farcs7 away from
CFRS14.0887,
   the FWKK radio map shows that there is a radio lobe centered on
   it (but with S $<$ 16$\mu$Jy).

\noindent
15V37: There is a very faint (I$_{AB} \sim$ 23.5) galaxy
   coincident with this source (offset 0\farcs5) but no spectrum was obtained.

\noindent
15V39: Excellent positional agreement (0\farcs5) with CFRS\-14.0854, a galaxy
   at z = 0.992 which has a faint (I$_{AB} \sim$ 23.2) companion only 2\farcs1
away.

\noindent
15V40: The positional agreement (0\farcs8) with CFRS\-14.0820 is good.

\noindent
15V41: Although CFRS14.0818 (z = 0.899) is 3\farcs5 W of the catalogued
position of
   this radio source, the highest resolution map given by
   FWKK shows a source
   coincident with this galaxy, and they comment that the radio source is
   probably a blend of emission from several galaxies (with S $<$ 16$\mu$Jy).
 In addition,
   the spectrum shows strong \hbox{[O\,{\sc ii}]} emission and strong Balmer
absorption
   lines indicating on-going star formation.

\noindent
15V42: The radio source is located rather far away (2\farcs5) from the centre
of
   CFRS14.0727, a large galaxy at z = 0.463, and hence may not be
   associated with it at all. However, the spectrum
   of this galaxy shows strong \hbox{[O\,{\sc ii}]} and \hbox{[O\,{\sc iii}]}
emission indicating
   that it probably is the radio source.

\noindent
15V45: There is a very faint (I$_{AB} \sim$ 24.3) galaxy coincident with this
radio source, but it is not visible on our $V$ image.

\noindent
15V47: A spectrum of the brightest part of CFRS\-14.0665, the nearest
 ($\sim$ 2 arcsec away)
   galaxy (or galaxies?) to the radio source was obtained, but no features
   were unambiguously identified, possibly indicating a high redshift.

\noindent
15V48: Excellent positional agreement (0\farcs2) with CFRS\-14.0663.

\noindent
15V49: Although the catalogued position for this radio source is 1\farcs9 away
   from CFRS14.0667, the highest resolution radio map shows a source coincident
   with it. Unfortunately, no spectrum was obtained, but the extended
   appearance of the galaxy suggests that z $\la$ 0.5 rather than z $\ga$ 1.

\noindent
15V50: Excellent positional agreement (0\farcs4) with CFRS\-14.0645
 but no spectrum was obtained.

\noindent
15V51: Excellent positional agreement (0\farcs4) with a faint red galaxy.

\noindent
15V53: Good positional agreement (1\farcs1) with CFRS\-14.0612 but no certain
features
   were identifiable on our rather poor spectrum.

\noindent
15V56: This source lies within the visible extent of a bright nearby spiral
   (CFRS14.0573; radio source 57). No distinct object is visible on
   our images at the location of the source, but since it is effectively
  masked by the bright foreground galaxy, it was removed from our
 complete sample.

\noindent
15V57: Excellent positional agreement (0\farcs6) with CFRS\-14.0573, a bright
   z = 0.010 galaxy displaying a strong starburst spectrum.

\noindent
15V59: This source is coincident with a very faint galaxy. The bright object
3\farcs9 E is an M star.

\noindent
15V60: Excellent positional agreement (0\farcs3) with CFRS\-14.9154.

\noindent
15V62: Nothing visible on any of our images (even summed)
 at the location of the source.

\noindent
15V65: Despite the apparent discrepancy (1\farcs4) between the position of
   CFRS14.0426 and the catalogued radio position, the highest resolution
   map published by FWKK shows a source coincident with
   CFRS14.0426, an M star.

\noindent
15V67: No optical counterpart of this compact source is visible on our
   I image, but a very faint object is present (at V $\sim$ 24.7)
   on the sum of our B, V, and I images. Unfortunately, we do not have a
K image.

\noindent
15V69: Excellent positional agreement (0\farcs2) with CFRS\-14.0393, but
 FWKK comment that the core associated with this galaxy is 13$\mu$Jy and that
   there is an extension to the E responsible for the remaining 10$\mu$Jy.

\noindent
15V70: Nothing visible on any of our images, including the sum of B, V and I,
at the location of the source. No K image was taken.

\noindent
15V72: The core (32$\mu$Jy according to FWKK) is apparently coincident
   with the faint galaxy located $\sim$1\farcs5 SW of the cross shown
   in Fig. 1 and is presumably identified with it. The bright object
   4\farcs3 E of the quoted radio source position, CFRS14.0274, is an M star.

\noindent
15V73: Good positional agreement (0\farcs8) with CFRS\-14.0276,
 one of the few optical counterparts
   which does not have \hbox{[O\,{\sc ii}]} emission.

\noindent
15V81: Good positional agreement (0\farcs7) with CFRS\-14.0154, a galaxy for
   which we derive z = 1.158 based on one strong emission line assumed
   to be 3727 \hbox{[O\,{\sc ii}]}.

\begin{figure}
 \vspace{10cm}
 \caption{Finding charts for all the 40 FWKK sources identified in the CFRS MOS
field. The areas shown are 20\farcs7 $\times$20\farcs7 with N at the top and E
to the left, and they were extracted from a combined B, V and I image
(except for 15V18 which was identified on a K image). The radio positions are
marked either by a cross or by a circle.}
 \label{sample-figure}
\end{figure}

\section {NATURE OF THE SOURCES WITH I$_{AB}$ $<$ 22.5}

\subsection{Morphological and photometric properties}

Most of the sources show typical morphologies for galaxies, i.e., round,
edge-on disk, or irregular shapes.  Only two sources (the quasar 15V12 and
the M star 15V65) are not
spatially resolved, while some sources show compact cores.
Deconvolution demonstrates that three sources (15V23, 47 and 49)
 present very complex
morphologies with more than three individual components within
2 arcsec (see Fig. 2). These components appear
to have very different colours one from another, and it is difficult to
understand the nature of these peculiar sources from our data.

\begin{figure}
 \vspace{5.5cm}
 \caption{Morphologies of some optical counterparts before and after
 deconvolution
 for (see text):
a) 15V23, I; b) 15V47, I; c) 15V49, I; d) 15V 26a and b; e) 15V18, K.}
 \label{sample-figure}
\end{figure}

 Histograms of the V and I magnitude
distributions are shown in Fig. 3
 for all sources of the complete sample, assuming that the two
unidentified sources have V$_{AB}>$25 and I$_{AB}>$24 respectively. The V
histogram
(Fig. 3a) shows the expected increase of the source number with
magnitude, until the magnitude limit is reached.
However, the I histogram (Fig. 3b) has a well-defined peak at I$_{AB}$ = 21.9,
much brighter than our completeness limit.

\begin{figure}
 \vspace{5.5cm}
 \caption{Histogram of the $\mu$Jy radiosource optical counterparts in a) V
band;
b) I band. A limit of V=26 and I=25 have been adopted for the undetected
sources.}
 \label{sample-figure}
\end{figure}

\subsection{Redshifts}

The spectra of all objects which we observed are shown in Fig. 4.
Secure redshifts were determined
for 19 of the counterparts,
and less reliable redshifts for two more (15V10 and 15V15), but
no redshift could be determined for 15V47.
Spectroscopic observations were also made
of 4 of the counterparts which we subsequently rejected because
their de-blended radio flux was too low, and a spectrum was obtained
of the very faint counterpart of 15V53, but
we were unable to determine its redshift.
 The completeness of our redshift
determinations of objects with S$>$16$\mu$Jy and I$_{AB}\le$ 22.5
is 86 per cent,
identical to that for the CFRS survey as a whole.

\begin{figure}
 \vspace{5.5cm}
 \caption{Spectra of 19 $\mu$Jy radiosource counterparts which resulted in
 secure redshift identifications.}
 \label{sample-figure}
\end{figure}

The main spectroscopic properties of the sources
are listed in Table 1.
In the double-limited sample, there is one M star (15V65), one QSO (15V12)
and one overluminous ($\sim$10L$^{*}$) galaxy  at
z = 1.149 (15V23) which was already noted to have a very complex morphology.
The rest of the sources can be divided into three different classes from
their colour and spectral properties and include 5 very blue emission-line
galaxies (26 per cent of the whole sample),
 6 spirals (32 per cent) and 5 ellipticals (26 per cent).

\subsection{Optical spectra and colours}

\begin{figure}
 \vspace{5.5cm}
 \caption{a) Hubble diagram for the spectroscopically identified sources.
Curves are
 (k-corrected) redshift magnitude relations for L* galaxies (early types to
later type), derived from Bruzual and Charlot (1993) models. Symbols are
filled circles for early type galaxies (E/S0), filled
squares for later type galaxies ($\sim$Sa to Sbc), filled triangles for
emission-line galaxies and stars for other peculiar objects.
b) V-I versus redshift diagram. Symbols and curves have the same meaning as
in Fig. 5a).}
 \label{sample-figure}
\end{figure}

The 5 emission-line galaxies (15V26a, 26b, 42, 57 and 81) have generally
blue V-I colours and are distributed over a large range in
redshift and absolute magnitude (Figures 5a and b).
At first sight these spectra resemble those of starburst galaxies which
Benn et al. (1993) claim to be common at 0.1 mJy levels. However,
detailed analyses of the spectra suggest that their
ionisation sources may be more closely related to those in AGNs.
We have quantified the emission line activity in these galaxies by
measuring their line ratios and plotting them in  the diagnostic diagrams
developed by Veilleux and Osterbrook (1987). 15V57 has
log(\hbox{[S\,{\sc ii}]}/\hbox{H\,$\alpha$}) = -0.41
and log(\hbox{[O\,{\sc iii}]} 500.7~nm/\hbox{H\,$\beta$})
= 0.83, which together with the presence of
\hbox{[O\,{\sc i}]}630.0 definitely indicates this object is
 a Seyfert 2 galaxy (Fig. 4).
This object has the lowest redshift of our sample (z = 0.01) and would be
classified as a dwarf blue elliptical from its luminosity, colour and
morphology. 15V26b is not as blue as the other ones (its V-I colour
suggests a Sa galaxy), but it has an emission line spectrum
characteristic of a LINER
(log(\hbox{[S\,{\sc ii}]}/\hbox{H\,$\alpha$}) = -0.32,
log\hbox{[N\,{\sc ii}]}/\hbox{H\,$\alpha$}= -0.26, with no \hbox{[O\,{\sc
iii}]}).
It is located in a large edge-on disk (Fig. 2d). 15V26a and 15V42 have
both \hbox{[O\,{\sc ii}]}3727, \hbox{[O\,{\sc iii}]}5007 and \hbox{H\,$\beta$},
which also allows
classification of
their ionisation source (see Tresse et al. 1994). The line ratios are
log\hbox{[O\,{\sc iii}]}/\hbox{H\,$\beta$} = 0.72$\pm$0.4, 0.58$\pm$0.2 and
log\hbox{[O\,{\sc ii}]}/\hbox{H\,$\beta$} = 0.95$\pm$0.3, 0.76$\pm$0.2
respectively for 15V26a and 15V42. Although the errors are
relatively large, to provide such intensity ratios requires high temperature
ionisation sources (T$\sim$100~000K), more characteristic
 of an AGN than massive
stars. Furthermore, we have not been able to estimate the extinction,
 so the calculated  \hbox{[O\,{\sc ii}]}/\hbox{H\,$\beta$} ratios are
underestimates.

15V26a has an
irregular morphology (Figure 2d), while 15V42 is a nearly edge-on disk
galaxy, and both have luminosities substantially lower than L$^{*}$.
The source with the highest redshift among these objects
 is a very blue emission-line galaxy
(15V81, z=1.158) which is brighter than L$^{*}$. We have no real indication
of the nature of this object
 since most of the emission lines are
outside our spectroscopic window.

Six galaxies (11, 19, 21, 24, 48, 60) have spiral-like characteristics,
with colours ranging from Sa to Sb,
 and disk-like morphologies. They are rather luminous galaxies
 (L$\sim$1.5$\pm$0.4L$^{*}$) and lie at
moderately high redshift (from z = 0.37 to z = 0.81, average z = 0.6). They all
show  moderate \hbox{[O\,{\sc ii}]} emission (average W =12\AA\ at rest), and
relatively
strong Balmer continuum and absorption lines. These features
are easily seen in the combined
spectrum (Fig. 6a). Equivalent widths of the Balmer absorption lines
range from 3 to 5\AA, indicating the presence of A and F stars
suggesting that strong star formation occurred in these objects
$\sim$1 Gyr ago. In addition, strong \hbox{[O\,{\sc ii}]} emission is present
in the spectra of these galaxies and so
we classify them as `S+A' galaxies by
analogy with the E+A galaxies described by Dressler \& Gunn (1983).

\begin{figure}
 \vspace{5.5cm}
 \caption{a) Summed spectra of  6 galaxies classified as S+A (poststarburst)
on the basis of their spectral features, colours and morphologies. Note the
strong Balmer absorptions and continuum which almost hide the 4000A break.
b) Combined spectra of the 6 galaxies classified as early-type galaxies.}
 \label{sample-figure}
\end{figure}

The spectra of five of the galaxies (28, 34, 39, 40, 73) are more
typical of ellipticals, with well-defined 4000\AA\ breaks and faint or no
\hbox{[O\,{\sc ii}]} emission lines (W = 3.5\AA\ at rest from the combined
spectrum, Fig. 6b). Their V-I colours are bluer than old and
non-evolved stellar populations
by $\sim$0.7 magnitude. They are luminous galaxies
(L $\sim$ 1.9$\pm$0.2 L$^{*}$) and lie at
higher redshifts  (0.75 to 0.99) than the `S+A' galaxies.

\subsection{The radio spectral index--colour diagram}

Figure 7 shows  the radio spectral index $\alpha$
between 1.5 and 5GHz for each identified object (filled points)
in our sample, versus their V-I colour.
 Each of the three $\mu$Jy populations occupy distinct
areas in the diagram: all red ellipticals have  inverted radio spectra
 ($\alpha$ = -0.4$\pm$0.3), all the S+As have  moderately steep
spectra ($\alpha$ = 0.40$\pm$0.18), while
the bluest emission-line galaxies have inverted spectra,
with the notable
exception of the most distant one, 15V81 ($\alpha$ = 0.7).
Inverted-spectrum radio emission from ellipticals  has been observed
in some nearby ellipticals (Wrobel and Heeschen 1984). According
to Rees (1984), this may indicate the presence of a low-power AGN,
although other alternatives are possible. It is unlikely that
the observed inverted radio spectra can be attributed to
opacities at low frequencies
since it only affects emissions below 1GHz at rest (see Condon 1992), i.e.,
below 0.5 GHz at z $\sim$ 1.

\begin{figure}
 \vspace{8cm}
 \caption{The radio spectral index versus V - I colour for all objects in the
complete sample. The symbols are the same as in Figure 5, except that open
circles
represent the faintest optical counterparts without spectral classification.
The
FWKK number of each source is given beside each symbol, and the arrows denote
limits.
Note the location of the early-type galaxies in the
bottom right, the S+As (post starburst) in the middle top and the emission-line
objects either in the top left (one starburst, 15V 81) or in the bottom left
(AGN induced emission-line galaxies). Note that the low V-I values for the
latter objects are also due to their moderate redshift. Delimiting areas are
indicated
from the location of spectroscopically-identified objects,
 separating the early type
galaxies (E/S0) from the post starburst galaxies (S+A). An extension of
the E/S0 area is also indicated for large redshift where V-I is likely to
decrease.
 A typical error bar is shown at lower left.}
 \label{sample-figure}
\end{figure}

Inverted radio spectra are also exhibited by the four very blue galaxies
at low and moderate redshift, supporting the hypothesis that AGNs
are present in their cores too, producing both the radio
emission and the emission lines. Note also that the QSO (15V12) lies
in the area of blue emission-line sources in Fig. 7.

The post-starburst spirals have radio slopes
noticeably flatter
 than the mJy starburst galaxies, which might indicate an
increasing contribution of thermal radiation from remnant supernovae
(Condon 1992).
 The radio emission from 15V81 could be ascribed to
star formation similar to that from starbursts observed at higher radio fluxes.
 Finally, the source 15V23 is more difficult
to interpret, since it has a positive radio index and there is some
uncertainty about its true V - I colour because of its complex optical
morphology.

\subsection{I$_{AB}<$22.5 objects with no redshifts}

Among the 25 I$_{AB}<$22.5 optical counterparts, we failed
to get secure redshifts for 3 of them, while for observational reasons we
were not able to observe three others.
There is some indication that the three spectroscopic `failures'
(15V10, 15 and 47) all lie at z $>$ 1.
In the V - I -- spectral index diagram, 15V10 lies
 in the area defined by ellipticals with low power AGN,
while 15V15 is slightly offset to bluer colour, as
expected if it is at z $>$ 1. 15V47 has a
very complex morphological type which leaves some uncertainty in the colour
of its optical counterpart; it may be similar to the high z object 15V23.
Hence it is possible that we failed to get good redshifts
for these objects simply because their
dominant spectral
features (\hbox{[O\,{\sc ii}]} 372.7 nm and 400.0~nm break) are beyond our
spectroscopic
window ($>$8500A), resulting in featureless spectra (Fig. 4).

The three objects for which we do not have spectra (15V5, 49 and 50)
all have relatively steep radio
spectra.  15V5 is likely to be an S+A galaxy from
its location in Fig. 7 and from its elongated morphology. 15V49
has the most complex morphology of our sample, while the nature of
15V50 is not clear, since it has a red colour and a steep radio slope.

\section {NATURE OF THE I$_{AB}>$22.5 COUNTERPARTS}

 The fact that we have redshifts and colours and hence have
been able to determine the nature of most of the sources with I$_{AB} <$
22.5, allows us to extend our analysis to the fainter sources, using
only their colours and spectral indices.
Among the 11 I$_{AB} >$22.5 sources,  two have
 no optical counterparts and
one is only detectable in the combined B+V+I  image, so it is impossible
to speculate on the nature of these.

\begin{figure}
 \vspace{5.5cm}
 \caption{V-I versus I-K colour indices. Typical evolutionary
tracks are shown for three classes of galaxies (solid lines), with three
fiducial
redshifts marked by dashed lines. As before, spectroscopically-classified
objects
are represented by filled symbols, objects without redshifts by open symbols.
There are three faint objects which show definite high I-K colours,
which are likely to be early-type galaxies at z $>$ 1.}
 \label{sample-figure}
\end{figure}

 There is considerable evidence that virtually all of the remaining
 sources are
likely to be at z $>$ 1.
Four of them, 15V33, 37, 45 and 72, have properties
similar to the high redshift early-type galaxies.
In fact, the two for which we have K photometry have
colours much redder
(I-K = 3.4 and 3.2 for 15V37 and 15V72 respectively) than any object in the
CFRS
 sample ($I_{AB} \le$ 22.5).
 This, their round morphologies,
and their V-I colours (Fig. 8) are all consistent with early-type galaxies at z
$>$ 1.
The reddest source of our sample is 15V18, with I-K$>$4.2. Its
morphology is surprisingly elongated, similar to that of an
edge-on disk (Figure 2). If this is the case, this source might be
a late-type galaxy at redshift even larger than 2.
The location of 15V53 in Fig. 7 supports
identification with a S+A galaxy, very probably also
at z $>$ 1, from its faintness and its colour.
15V59 is located near the QSO (15V12) in Fig. 7,
and hence might be either a QSO or an AGN-powered source.
After deconvolution, 15V59 appears to be composed
of two compact components.
15V51, which has an elongated morphology, appears to be similar to
15V50 and is also a  candidate to be a radio galaxy at very high redshift.

\section{THE MICROJY RADIO SOURCES}

Combining the results from the two previous sections,
an estimate can be made of the relative contribution
of the different types of objects to the $\mu$Jy
population. We then find, among the 36 FWKK radio sources with
S$>$16$\mu$Jy in our MOS field:

\begin{description}
  \item 5 definite and 6 probable early-type galaxies ($\sim$33 per cent of
sample).
The latter 6 are probably at z$>$1.
   \item 7 S+A's (21 per cent) including 1 possibly at z$>$1.
  \item 7 blue emission line objects (21 per cent), including 1 (or 2) QSOs.
  \item 3 sources (9 per cent) with a very complex optical morphology,
including
 one at z=1.15.
  \item 2  candidates for very high redshift radiogalaxies.
  \item 1 very elongated object with colours consistent with location
at very high z.
  \item 1 starburst at z = 1.135.
  \item 1  M star.
  \item 3 undetected or barely detected objects.
 \end{description}

The microJy population is hence
mainly constituted of three distinct populations
of galaxies with different redshift regimes: early-type galaxies
at z $>$ 0.75,
S+A at intermediate redshifts (z = 0.375 to z = 0.8 or slightly $>$ 1),
 and emission-line galaxies  at z $<$ 0.45 containing AGNs.

It is possible that the early-type galaxies may be over-represented,
since three of the spectroscopic identifications belong to the
structure at z = 0.985 found in this field (Le F\`evre et al. 1994), and it is
not
clear whether this has any effect or not.
 The fraction of $\mu$Jy sources with z $>$ 1
is  38 - 42 per cent,
depending on whether one assumes the unidentified sources are all
at high z, or if they have the same distribution as the rest of the sample.

\section{DISCUSSION}

\subsection{Comparison of $\mu$Jy with sub-mJy sources}

 At bright flux levels near $\sim$ 1 mJy, the radio source population is known
to be
composed largely of starburst galaxies at moderate redshifts.
However, even though it is believed that this may continue to
the sub-mJy levels, no
spectroscopic sample as complete as the one presented here has
yet been gathered. Thuan and
Condon (1987) have shown that optical-infrared colours of sub-mJy
radiosources are similar to that expected for Irr or starbursting
galaxies, and Rowan-Robinson et al. (1993)
have discussed their evolution, but conclusions
are limited by the low identification rate (112 of 523) of the sources.
Benn et al. (1993) were only able to identify
the brightest (B$<$22.3) optical counterparts and may have
missed most of the z$>$0.5 identifications (see Colless et al. 1990).
Windhorst, Dressler and Koo (1987) also obtained spectra for a small sample
of only $\sim$10 sub-mJy radiosources, all of which show \hbox{[O\,{\sc iii}]}
and \hbox{H\,$\beta$}
 lines with W $\sim$20-30\AA.

Our data indicate that the star formation activity, as well
as the corresponding radio power, may decrease towards lower flux limits.
Only one galaxy in our sample has a classical starburst spectrum,
 and it is
at a much higher redshift than the Benn et al. sources.
As noted above, most of the
emission-line galaxies identified by us appear, from
their emission line ratios and from their inverted or flat radio spectra,
to be powered by AGN rather than starbursts. On the other hand,
the S+A  objects in our sample
may well be the remnants of an active starburst population.
Starburst activity is expected to last for only a relatively short period
(typically 10$^{8}$ yrs), followed by a  decrease in
radio and emission line activity which in turn produces a less steep
radio spectrum (increasing contribution
from the thermal radiation by supernovae remnants) and optical spectra
dominated by several Gyr old stars (A and F stars), with faint \hbox{[O\,{\sc
ii}]} and no
\hbox{[O\,{\sc iii}]} emission. The surface density of the S+A post-starburst
population
found at $\mu$Jy levels -- assuming the count slope of
1.18 from FWKK -- is indeed ten times larger than that of the mJy
population at 1.4GHz, which is assumed to be mainly composed of starbursting
galaxies. It is thus possible that the S+A galaxies are the remnants of
generations of active starburst galaxies.

\subsection{Radio emission from distant ellipticals compared to nearby ones}

The presence of radio emission in a large fraction of nearby early-type
galaxies  is well known, and has been widely studied by Wrobel and Heeschen
(1991) in a complete survey at 6cm.
They found that $\sim$30 per cent of nearby ellipticals have radio emission in
their cores at powers ranging from
 10$^{19}$ to 10$^{21}$ W Hz$^{-1}$, which are
believed to be linked with low-power AGNs, since the core radio emission of
a small subsample have flat or inverted radio spectra.
The radio emission from the high redshift early-type galaxies in our sample
probably has the same origin, since for all but one (15V10),
the radio emission is concentrated in regions smaller
than 0\farcs2 to 2\arcsec.
However, at z$>$0.75,
more than a third of the early-type galaxies have P $>$ 10$^{23}$ W Hz$^{-1}$,
much larger than those in the Wrobel
and Heeschen sample. The samples are not directly comparable, however,
 since  only 4 of their galaxies have comparable optical luminosity
to those in our sample, and their radio power is more than ten times lower.

It is thus more relevant to compare our results to higher flux limit surveys.
In the Parkes sample (Wall et al, 1971; Downes et al, 1986; Dunlop et al.
1989), at the flux limit of $S_{2.7GHz}=100 mJy$, no flat
spectrum radio galaxy was found below z = 0.02, the redshift at
 which a 10$^{23}$ W Hz$^{-1}$ source is at the Parkes flux limit.
 We find 5 flat spectrum ellipticals
(0.75 $<$ z $<$ 1) in the volume contained in our 10\arcmin $\times$ 10\arcmin
{}~MOS field, which is 7.5
times smaller than the volume in the Parkes survey to z = 0.02.
Dunlop et al. (1989) found one similar source at z = 0.118,
PKS 1215+013 (P=5 10$^{24}$ W Hz$^{-1}$ and $\alpha$ = -0.15), but again
the corresponding volume is more than 1000
times larger than our CFRS volume.  Note that the bulk of the
galaxy population sampled in the CFRS lies at z = 0.3 to 1.

 We thus conclude that radio flat spectrum sources (P $\sim$ 5
10$^{23}$ W Hz$^{-1}$) occur much more often in high redshift early-type
galaxies than in low redshift ones.
  The fact that the
 fraction of radio emitting
early-type galaxies in the $\mu$Jy population is increasing
with the redshift (0 per cent at
z$<$0.5, 25 per cent at z$<$1 and 40 per cent at z$>$1) also demonstrates
 that early-type galaxies are experiencing strong evolution of their radio
properties beyond z = 0.75.  Indeed, one third of them apparently
exhibit AGN activity at z $\sim$ 1. One cautionary note must be added however,
since
Le F\`evre et al. 1994 detected a large structure at z $\sim$1 in this field,
and
so it is conceivable that this field may be unusual.

\section{CONCLUSION}

During the course of the CFRS project we have
observed 36 $\mu$Jy radiosources which form a complete subset of the
FWKK sample (S$_{5GHz}$ $>$ 16$\mu$Jy).
Optical counterparts have been identified for 94 per cent of these,
and redshifts have been determined for 86 per cent of those with
I$_{AB} <$22.5. Among the latter, which are mainly at z$<$1,
three different populations of comparable weight (from 26 to 32 per cent)
are identified:
 emission-line galaxies whose activity resembles that in AGNs,
 luminous post-starburst spirals
containing many A or F stars, and high redshift luminous early-type galaxies
containing low-power AGNs.

Based on our knowledge of the redshifts, colours and radio spectral indices
of the brighter objects, we are able to extend our analysis to
all sources in the radio flux-limited sample.
We find that 40 per cent of the total sample
 $\mu$Jy radiosources probably lie at z $>$ 1,
and nearly half are early-type galaxies.

The fact that blue galaxies with emission lines in our sample
also have inverted radio spectra is additional evidence that low-power
AGNs may reside in their cores, in accord with the results at lower redshift
discussed by Tresse et al. 1994.
We also postulate that the S+A galaxies in our sample may be
the result of evolution of starbursting galaxies
to a more quiescent state, as indicated by both their
radio emission and optical emission lines, since their surface
density is consistent with that of mJy starburst sources.
The sources corresponding to luminous early-type galaxies, a population which
represents
an increasing fraction with redshift (40 per cent at z $>$ 1),
also frequently appear to contain a low-power
 AGN nuclear source.  Although Wrobel and Heeschen (1991) report
similar radio activity in ellipticals at low redshift, they have
much lower radio power.

 The mixture of the three dominant populations in the $\mu$Jy counts
explains the apparently
discrepant V and I histograms of their optical counterparts.
Red early-type galaxies at high redshift contribute to the I$_{AB}$ = 21.9
peak, but they have very faint V magnitudes.
The strong decrease of the radio spectral
index from sub-mJy to $\mu$Jy counts appears to be due
 to a combination of three
factors: (1) the emergence of an elliptical population at high redshifts with
moderate radio emission (2) an increasing fraction of
narrow emission-line AGNs (Seyfert 2 and LINER); (3) a higher
contribution of the thermal radiation to the radio emission from spirals, and
the almost complete disappearance of starburst galaxies.
The fact that radio sources have much flatter radio spectra
at $\mu$Jy levels compared to those above 0.1 mJy can thus
be mostly attributed to the emergence of radio sources driven by
low-power AGNs ($>$ 50 per cent of the whole $\mu$Jy population).
Since the space density of AGN-driven sources apparently overtakes
those powered by stellar emission,
the contribution of AGN light to the faint source counts should be
reevaluated. Finally, our results demonstrate that a similar
study of a complete sample at sub-mJy levels should be carried out,
as well as an extension of the present survey to
other fields surveyed at $\mu$Jy depth.

\section*{Acknowledgments}
We thank J. Kristian and R. Windhorst for discussions, the referee, J. Wall for
helpful comments and the directors
of the CFHT for their continuing support and encouragement.
SJL's research is supported by the NSERC of Canada.
We acknowledge some travel support from NATO.

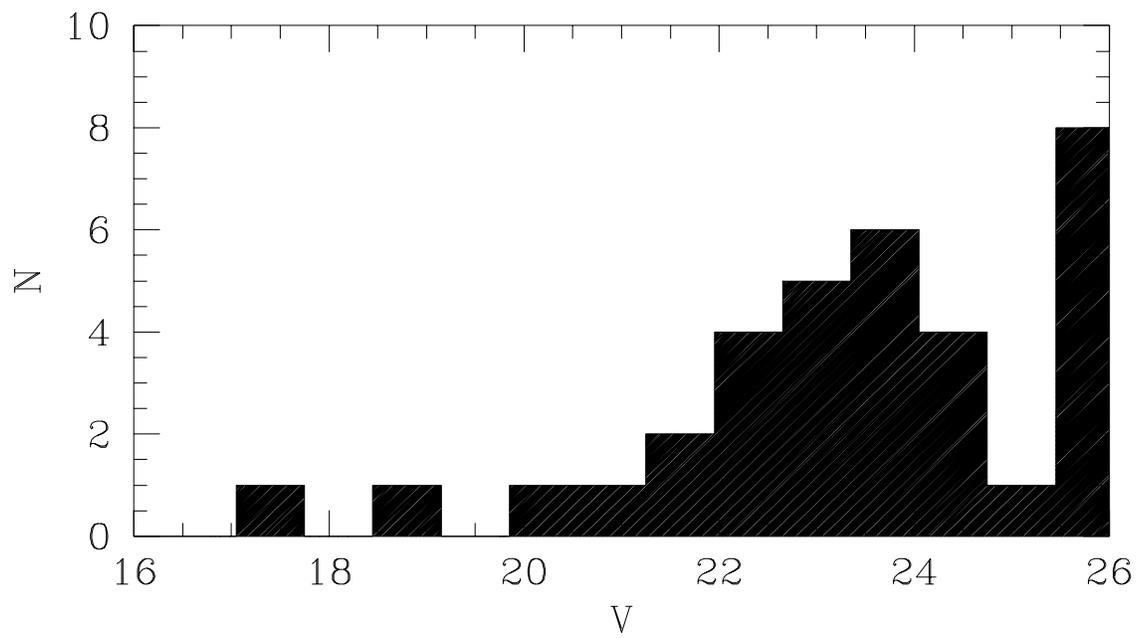
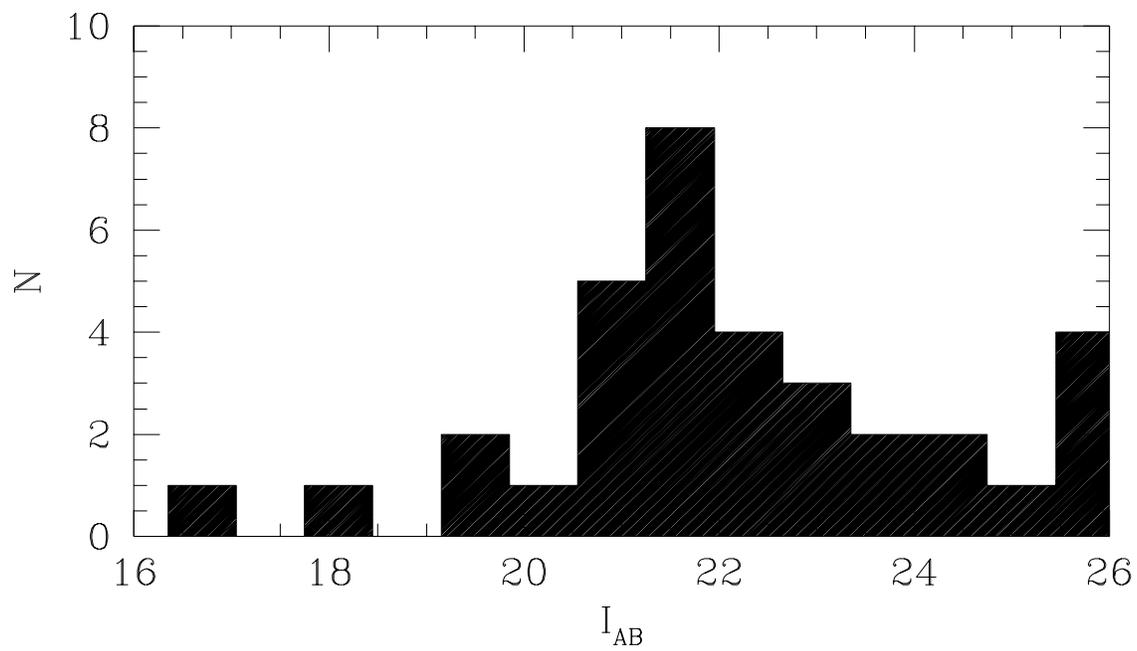
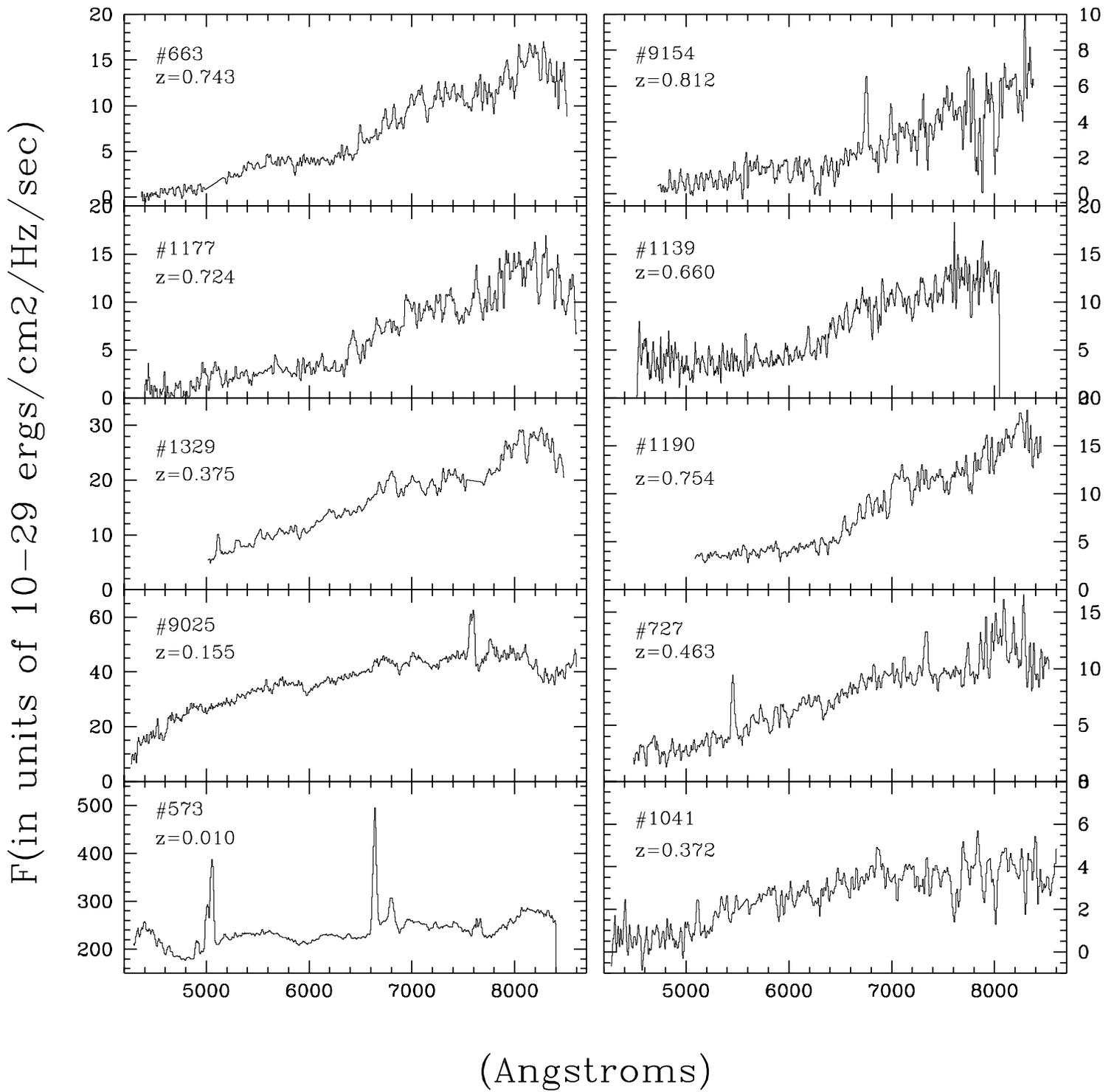
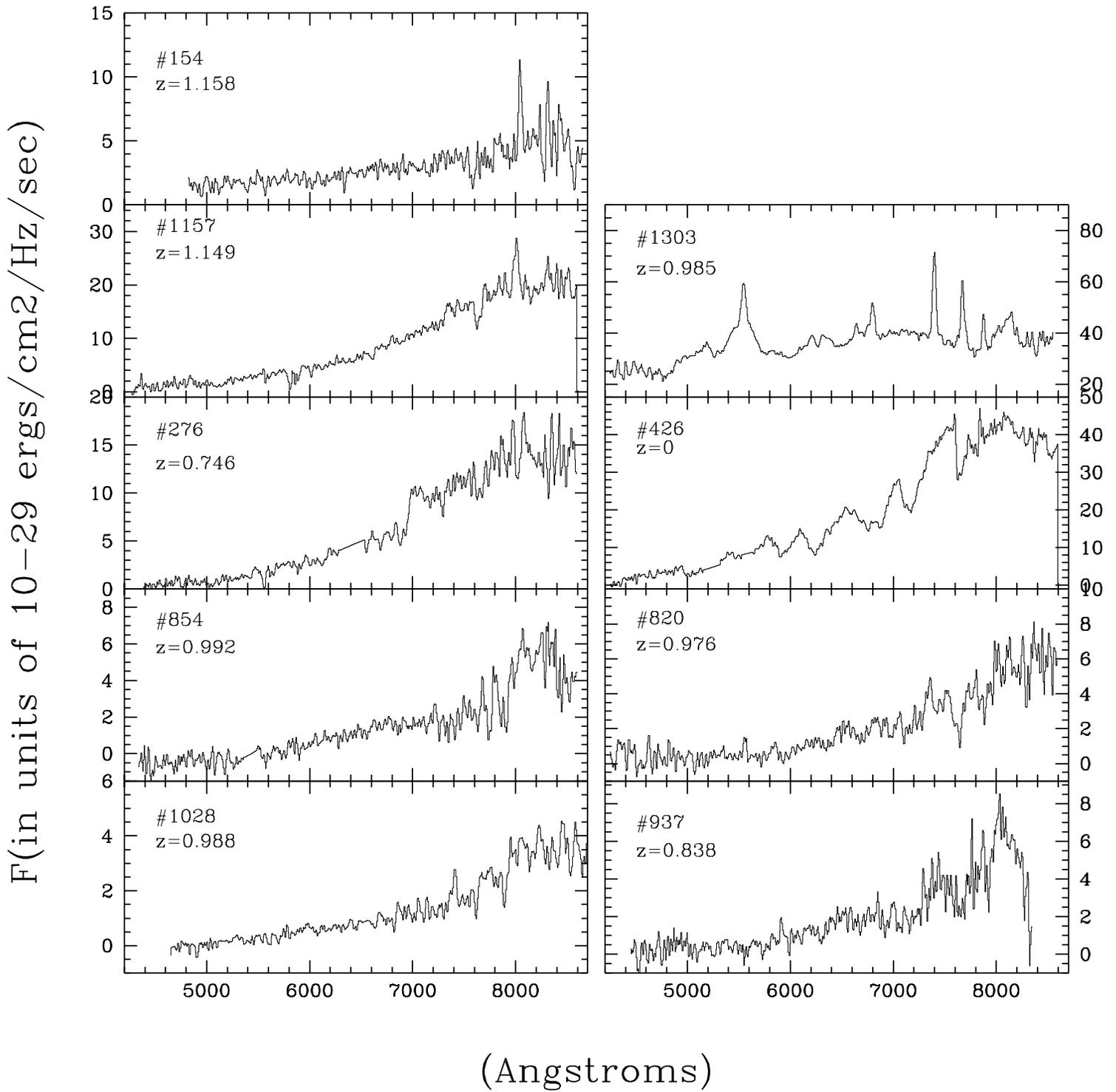
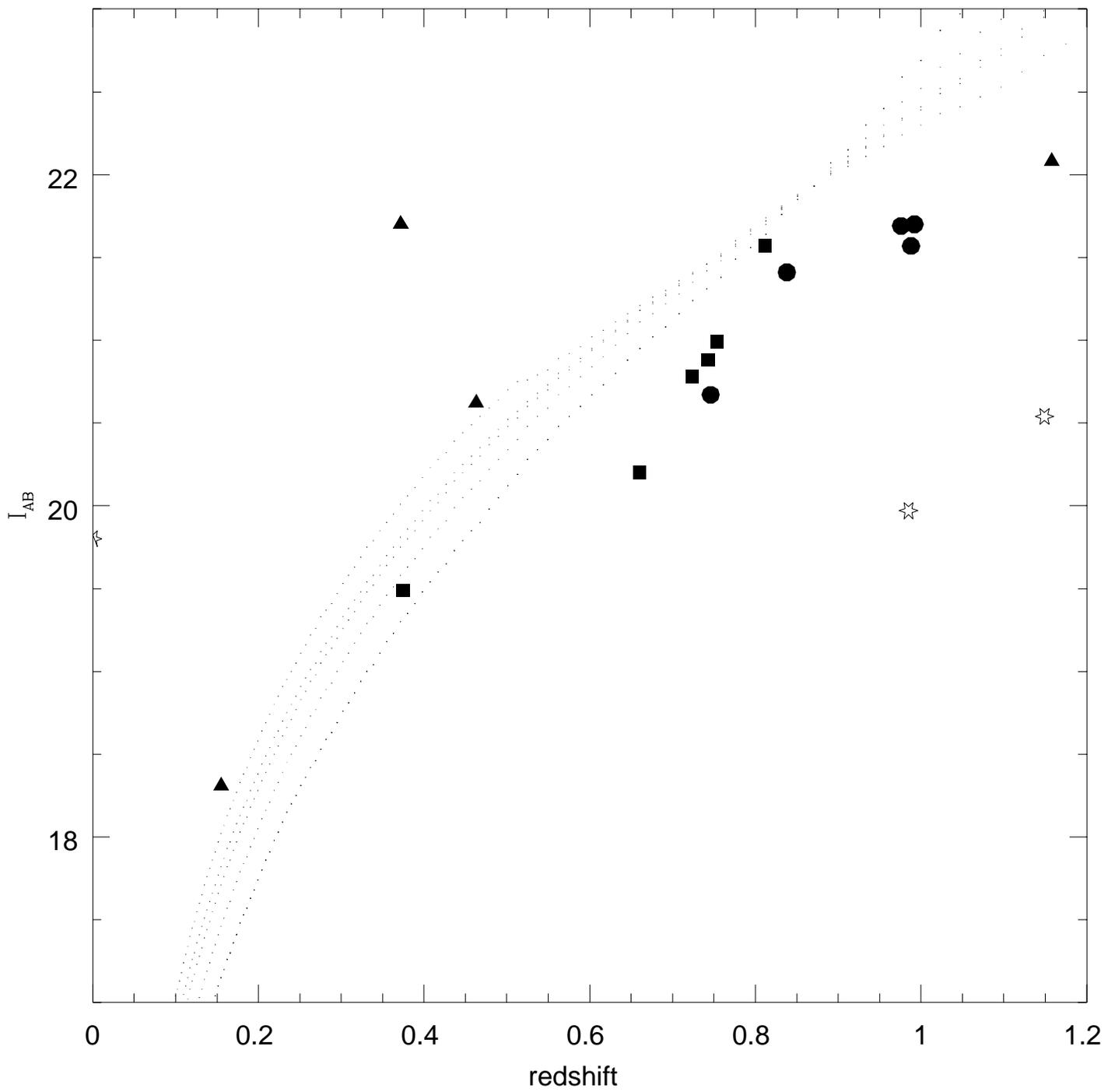
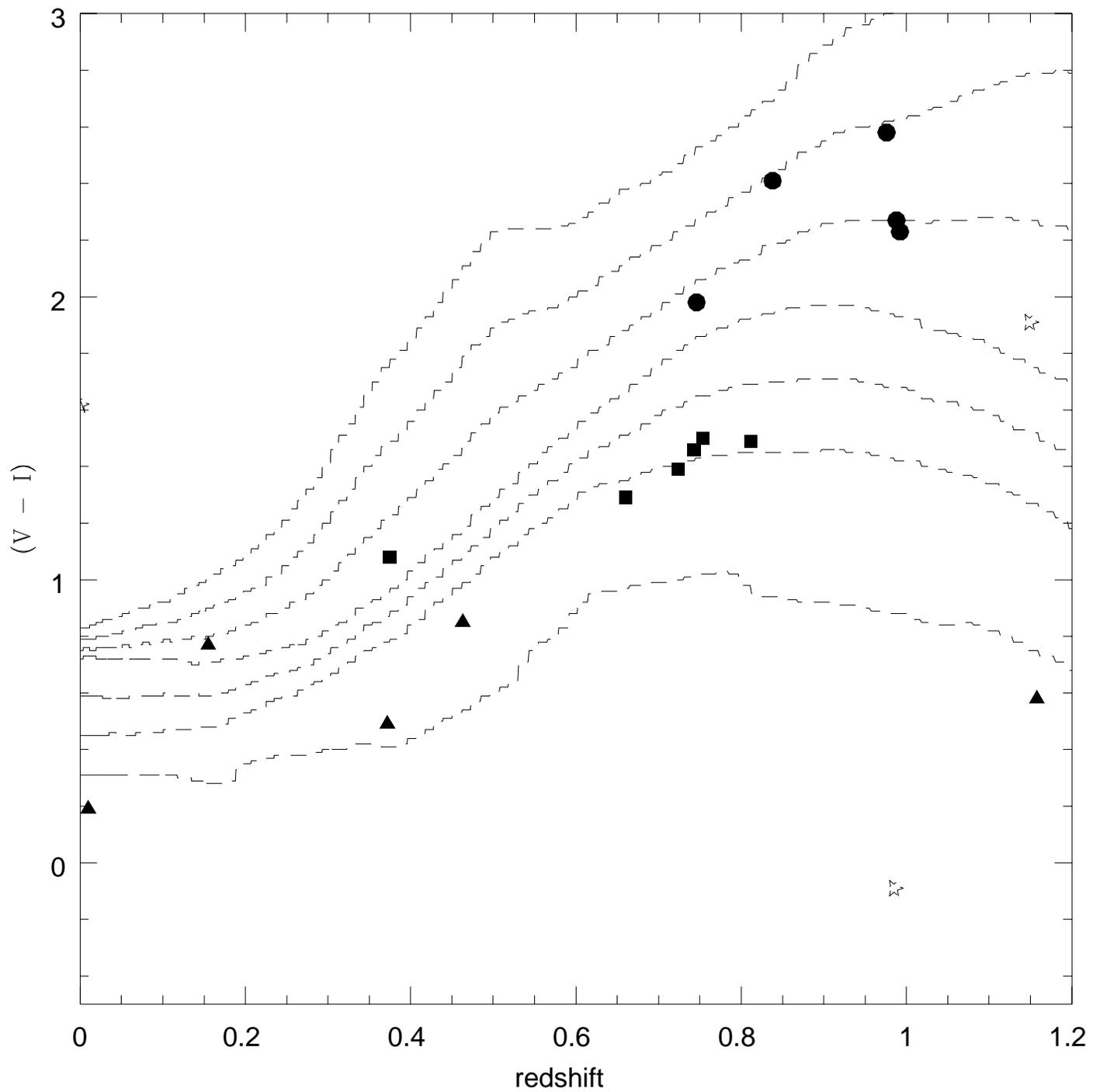
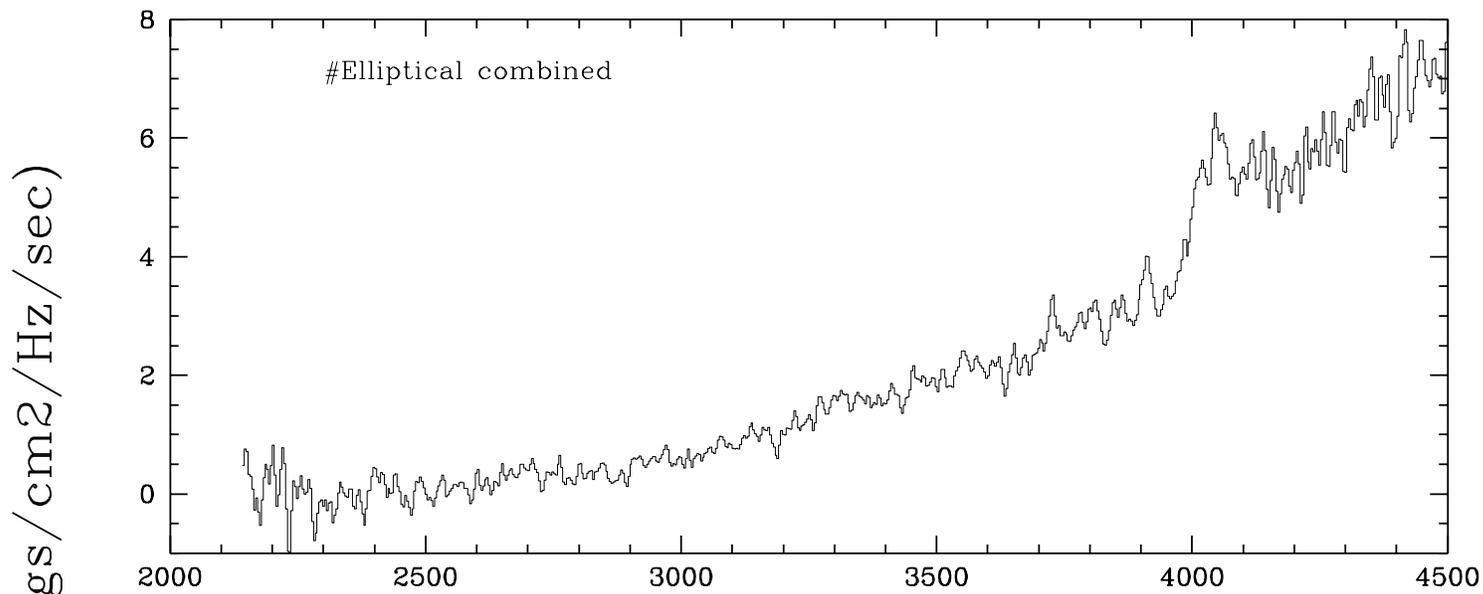
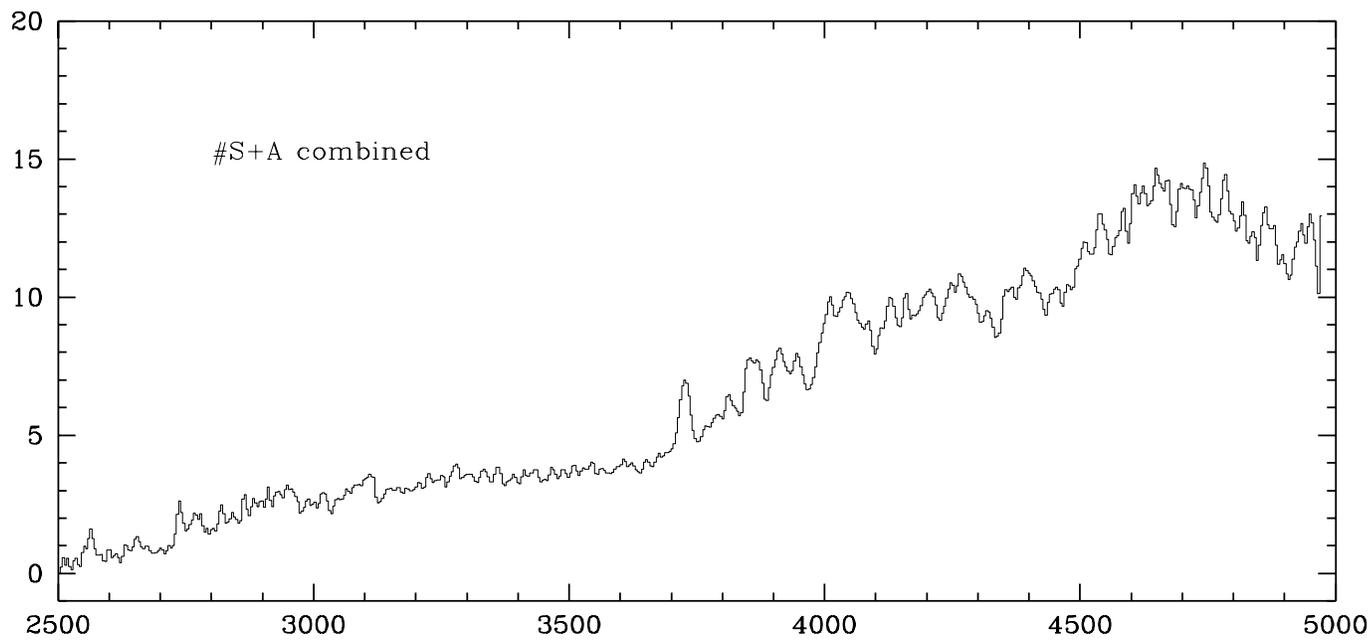
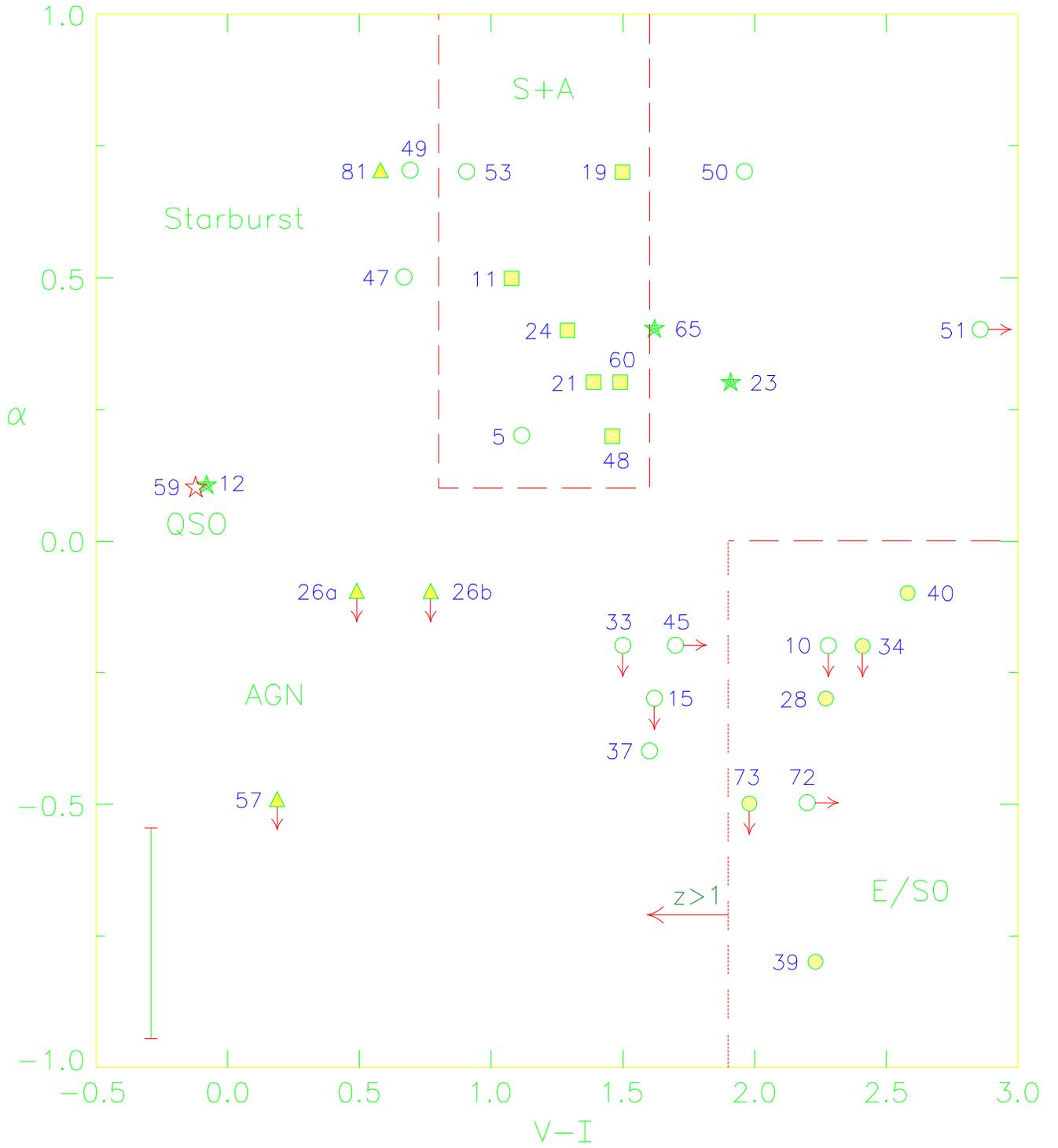
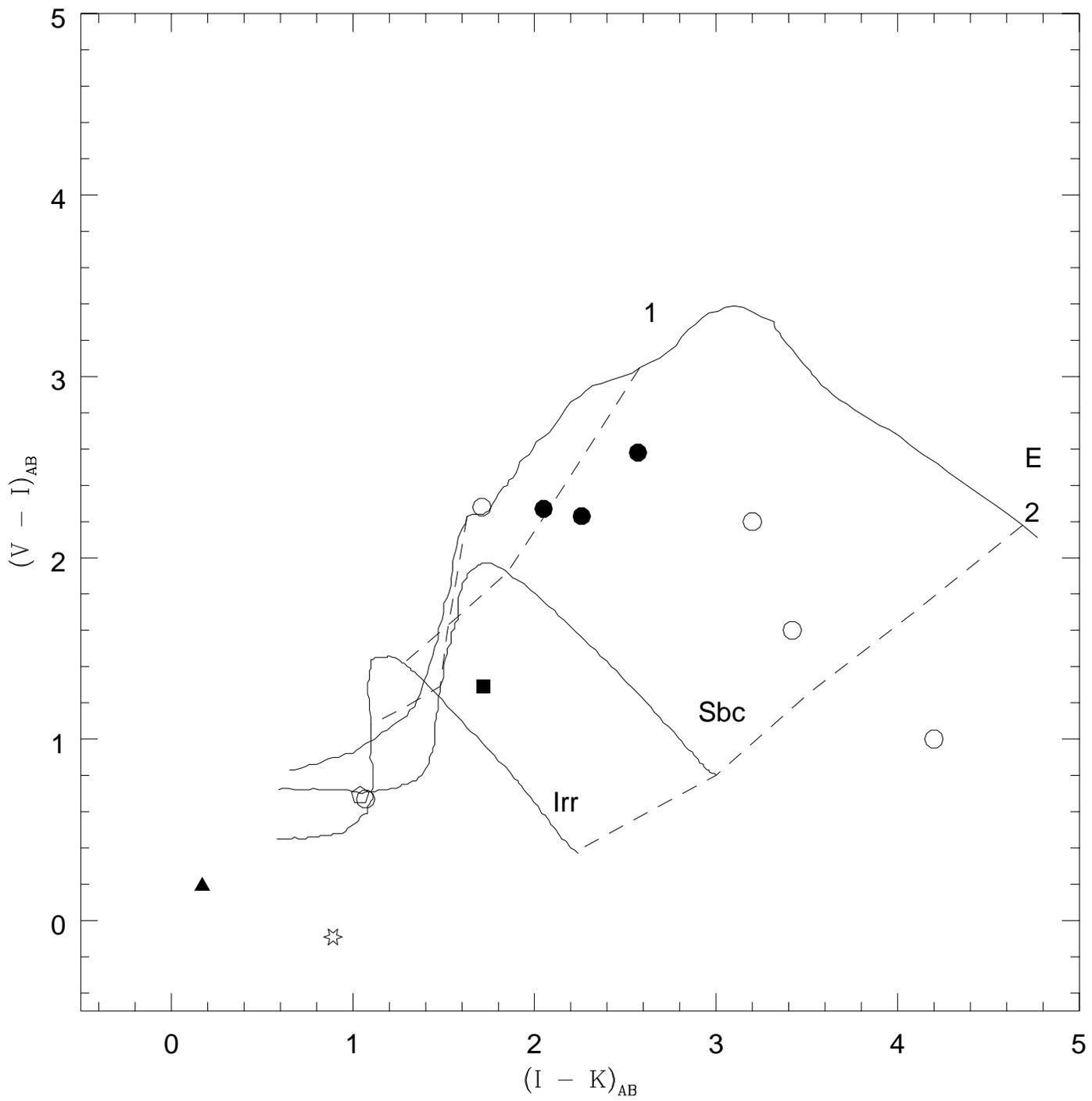
\end{document}